\documentclass[sigconf]{acmart}
\usepackage{subcaption}
\usepackage{multirow}
\AtBeginDocument{%
  }

\setcopyright{acmlicensed}
\copyrightyear{2018}
\acmYear{2018}
\acmDOI{XXXXXXX.XXXXXXX}
\acmConference[Conference acronym 'XX]{Make sure to enter the correct
  conference title from your rights confirmation email}{June 03--05,
  2018}{Woodstock, NY}
\acmISBN{978-1-4503-XXXX-X/2018/06}

\begin{document}

\title{Layer-wise Weight Selection for Power-Efficient Neural Network Acceleration}




\author{Jiaxun Fang}
\affiliation{%
 \institution{Shanghai Jiao Tong University}
 \city{Shanghai}
 \country{China}}
 \email{oxygenfunction@sjtu.edu.cn}

\author{Grace Li Zhang}
\affiliation{%
  \institution{Technical University of Darmstadt}
  \city{Darmstadt}
  \country{Germany}}
 \email{grace.zhang@tu-darmstadt.de}



\author{Shaoyi Huang}
\affiliation{%
  \institution{Stevens Institute of Technology}
  \city{Hoboken}
  \country{USA}}
\email{shuang59@stevens.edu}

\renewcommand{\shortauthors}{Trovato et al.}

\begin{abstract}

  Systolic array accelerators execute CNNs with energy dominated by the switching activity of multiply accumulate (MAC) units. Although prior work exploits weight dependent MAC power for compression, existing methods often use global activation models, coarse energy proxies, or layer-agnostic policies, which limits their effectiveness on real hardware. We propose an energy aware, layer-wise compression framework that explicitly leverages MAC and layer level energy characteristics. First, we build a layer-aware MAC energy model that combines per-layer activation statistics with an MSB-Hamming distance grouping of 22-bit partial sum transitions, and integrate it with a tile-level systolic mapping to estimate convolution-layer energy. On top of this model, we introduce an energy accuracy co-optimized weight selection algorithm within quantization aware training and an energy-prioritized layer-wise schedule that compresses high energy layers more aggressively under a global accuracy constraint. Experiments on different CNN models demonstrate up to 58.6\% energy reduction with 2-3\% accuracy drop, outperforming a state-of-the-art power-aware baseline.

\end{abstract}




\maketitle

\section{Introduction}
Convolutional neural networks (CNNs)~\cite{krizhevsky2012imagenet, simonyan2014very, he2016deep} have become the backbone of modern computer vision tasks and are widely deployed on both cloud servers and resource-constrained edge platforms. As the computational throughput of DNN accelerators continues to increase, energy efficiency has become a primary design constraint, especially in battery-powered or thermally limited environments. Systolic-array-based architectures~\cite{jouppi2017datacenter, chen2016eyeriss,9474215} , such as TPU-like matrix multipliers, dominate current hardware designs due to their high utilization and regular dataflow. In such architectures, the multiply-accumulate (MAC) unit is the fundamental computational block, and its switching activity strongly impacts total inference energy.

Recent studies~\cite{petri2023powerpruning} have shown that different weight values induce different levels of switching activity and delay inside MAC units, resulting in measurable variation in both power and timing behavior. These observations have motivated power-aware pruning and quantization techniques that bias network parameters toward lower-energy operations without changing the hardware. However, existing approaches still face several limitations when applied to full-scale CNNs on realistic accelerators.

First, power-aware pruning or quantization strategies often rely on oversimplified energy proxies~\cite{liu2017learning, gordon2018morphnet, he2018amc,10137171} instead of accurate MAC-level switching activity. These proxies correlate poorly with real hardware energy and thus limit the effectiveness of compression when deployed on systolic-array accelerators. Prior MAC-level power models~\cite{petri2023powerpruning}  are typically global: they measure average MAC energy over the entire network without accounting for layer-to-layer variation in activation distributions. Since activation statistics differ substantially across layers and depend on training dynamics and nonlinearities, global modeling fails to capture per-layer MAC behavior accurately, leading to suboptimal optimization decisions.

Second, many existing compression methods~\cite{molchanov2016pruning, han2015deep,10546870,10595861}  apply uniform pruning or weight restriction across layers. Such layer-agnostic schemes ignore the fact that different layers contribute very differently to overall energy consumption and have different sensitivities to perturbation. Treating all layers equally leads to inferior energy--accuracy trade-offs, especially when high-energy layers are not given priority.

To address these limitations, we propose a design framework that integrates layer-specific MAC energy modeling, an energy--accuracy co-optimized weight-restriction algorithm, and an energy-prioritized layer-wise compression schedule. By coupling hardware-level characterization with training-time optimization, the proposed method achieves substantially improved energy--accuracy trade-offs for CNN inference without any hardware modification.

Our main contributions are summarized as follows:
\vspace{-0.1in}
\begin{itemize}
  \item \textbf{Energy-prioritized layer-wise compression.}
  We propose a layer-wise compression algorithm that prioritizes layers according to their estimated energy contribution and processes them in a descending-energy order. For each layer, we explore a set of pruning ratios and restricted weight-set sizes, and select the most aggressive configuration that satisfies a global accuracy constraint. By focusing compression effort on high-energy layers and applying layer-specific strategies, our method achieves significantly better energy-accuracy trade-offs than uniform or global compression schemes under comparable overall compression ratios.

  \item \textbf{Layer-aware MAC and convolution-layer energy modeling.}
  We develop a per-layer MAC energy modeling framework that measures switching activity under layer-specific activation and partial-sum transition distributions. A compact MSB--Hamming-distance grouping scheme is introduced to approximate the large 22-bit partial-sum space, and the resulting MAC model is combined with an im2col-based, tile-level systolic mapping to estimate convolution-layer energy. This model delivers more accurate layer-wise energy estimates than prior global approaches and directly supports energy-prioritized compression decisions.

  \item \textbf{Energy-accuracy co-optimized weight restriction.}
  We integrate a co-optimized weight-selection algorithm into quantization-aware training~\cite{jacob2018quantization, krishnamoorthi2018quantizing}. Rather than naively choosing the lowest-energy weight values, we start from a safe candidate set and perform greedy backward elimination, jointly considering the estimated energy saving and the accuracy drop for each removal. This procedure constructs compact, expressive weight sets that effectively reduce MAC energy while preserving model performance, especially under aggressive weight constraints.
\end{itemize}

The remainder of the paper is organized as follows. Section~\ref{sec:motivation} discusses the motivation and problem setting. Section~\ref{sec:energy_modeling} presents the proposed energy modeling framework, and Section~\ref{sec:compression} details the layer-wise compression and weight-selection algorithms. Experimental results are reported in Section~\ref{sec:experiments}, and Section~\ref{sec:conclusion} concludes the paper.

\section{Motivation}
\label{sec:motivation}
\begin{figure}[h]
  \centering
  \includegraphics[width=0.80\linewidth]{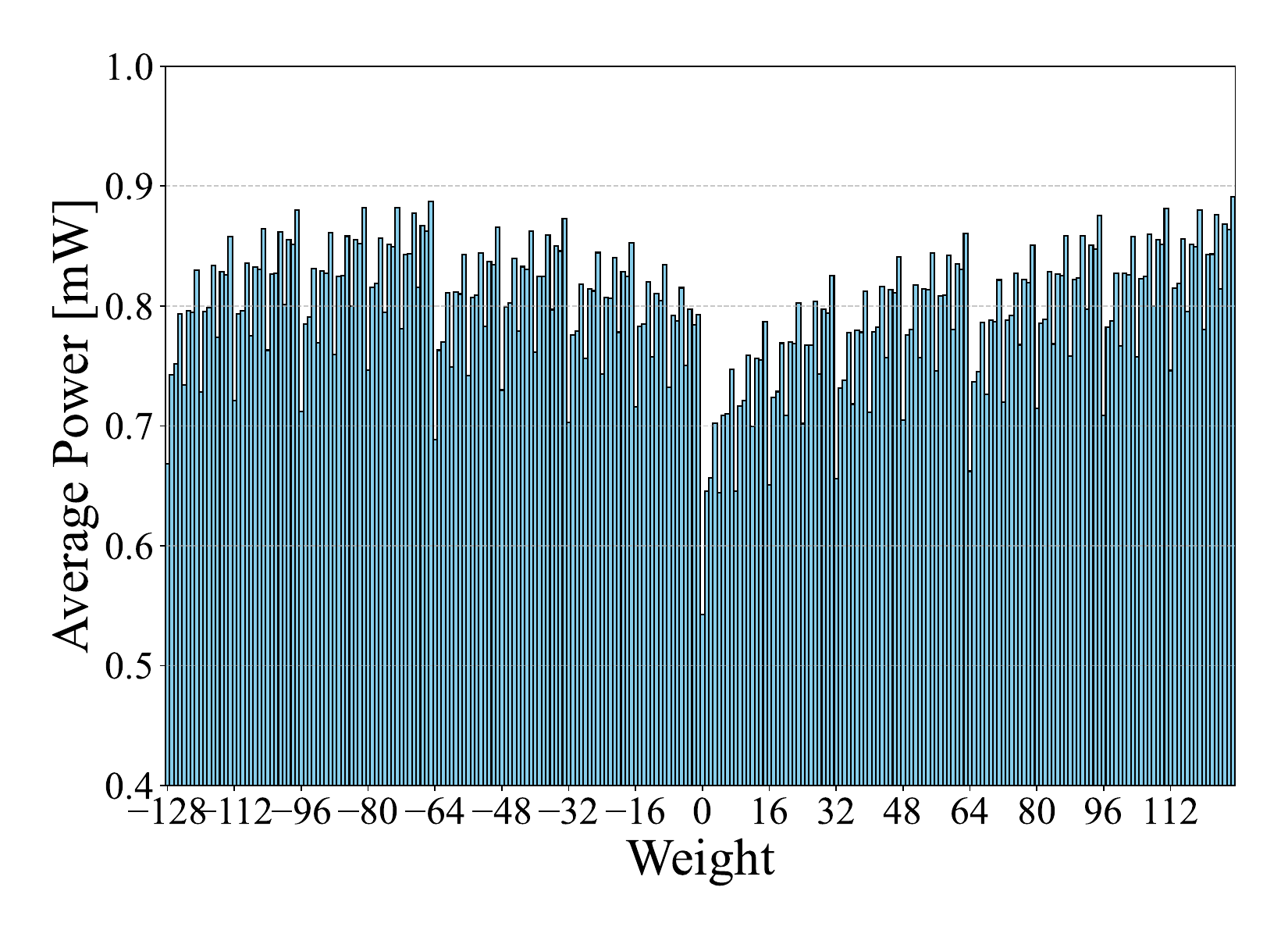}
  \caption{Average MAC power on different weight values}
  \Description{TO DO}
  \label{fig:weight}
\end{figure}

Energy consumption in systolic-array CNN accelerators is dominated by the switching activity of MAC units. Prior work such as PowerPruning~\cite{petri2023powerpruning} has shown that different weight values exhibit significantly different switching power, suggesting that restricting a model to low-energy weights can reduce MAC power during inference. Figure \ref{fig:weight} illustrates this variability by showing the average MAC power measured for different weight values in our simulation framework. This substantial spread confirms that weight selection is a promising direction for energy-aware model compression.

However, directly extending existing power-aware schemes to full CNNs exposes several limitations. A key challenge lies in how prior methods approximate activations. Many approaches~\cite{frankle2018lottery} assume a global activation distribution, implicitly treating all layers as statistically similar. In practice, activation sparsity, magnitude, and transition patterns vary greatly across layers due to architectural depth, nonlinearities, and dataflow differences. When an energy model ignores layer-specific statistics, its estimation becomes biased, which leads to suboptimal or even misleading compression decisions. Our results in later sections show that such misestimation limits achievable energy reduction and may degrade accuracy unnecessarily.

A second issue arises when selecting weight values purely according to their average MAC energy. Although low-energy weights reduce switching activity, choosing them without considering their representational importance weakens the model’s expressive capacity. We observe that naively selecting the lowest-power weights causes severe accuracy degradation, even when the number of allowed weights remains relatively large. This indicates that energy cost must be balanced with accuracy sensitivity during weight restriction. A more principled optimization—one that integrates both energy modeling and performance constraints—is required. We develop such a weight-selection mechanism in our method.

Finally, although layer-aware techniques have been explored in pruning and quantization works~\cite{liu2017learning,gordon2018morphnet,he2018amc}, they have not been applied to power-aware weight selection. This omission is critical: convolutional layers differ substantially in redundancy, activation volume, and energy contribution. CNN energy distribution is highly imbalanced, often dominated by a few early or high resolution layers. Applying uniform compression across layers therefore fails to exploit the potential for targeted energy reduction. A layer-wise strategy, informed by accurate per-layer energy modeling, can prioritize high energy layers and apply stronger restrictions only where the energy payoff is meaningful, while preserving information contained in more sensitive or low-energy layers.

These observations together motivate the need for a new compression framework that incorporates: (1) accurate, layer-specific energy modeling, (2) an energy–accuracy co-optimized weight selection mechanism, and (3) a layer-wise compression schedule that prioritizes energy critical layers.


\section{Energy Modeling}
\label{sec:energy_modeling}

This section introduces the proposed energy modeling framework, which characterizes the switching behavior of MAC units and estimates the energy consumption of full convolution layers executed on a $64\times 64$ weight-stationary systolic array. The framework consists of two components. First, we derive accurate per-weight MAC power by constructing probabilistic models of activation and partial-sum transitions. Second, we present a general formulation for estimating convolution-layer energy through a tile-based mapping of matrix multiplications onto the systolic array. These models form the foundation of the layer-wise energy-prioritized compression algorithm in Section~\ref{sec:compression}.

\subsection{MAC-Level Power Estimation}
We begin by estimating the average switching power of MAC units under realistic activation and partial-sum transitions. For each fixed weight value, we simulate the MAC cell within the $64 \times 64$ weight-stationary array and measure its switching activity across a variety of input transitions. Two practical challenges arise.
\textbf{\textit{Challenges 1}}: Partial-sum transitions occupy an enormous space: a 22-bit accumulator yields a theoretical transition space of size $2^{22} \times 2^{22}$, making direct statistical estimation impossible. Thus, we must approximate this distribution efficiently while preserving the key behaviors that influence dynamic power.
\textbf{\textit{Challenges 2}}: Activation statistics differ significantly across layers. In particular, layers with ReLU exhibit high sparsity, whereas layers using alternative nonlinearities or preceding pooling layers may produce denser activations. As a result, transitions cannot be modeled using a global distribution. To maintain accuracy, we collect activation and partial-sum statistics separately for each convolution layer.

\subsubsection{Partial-Sum Transition Grouping}

To model activation and partial-sum transitions from real data, we treat the two distributions independently. Activation transitions can be directly sampled from traced input sequences. However, the partial-sum distribution must be approximated.
We accomplish this by grouping partial-sum transitions so that (i) values within the same group exhibit similar switching behavior, and (ii) transitions drawn from the same pair of groups produce similar MAC power. This allows us to model the partial-sum space with a manageable number of 
representative 
clusters.

\begin{figure*}[htbp]
  \centering 
  \begin{subfigure}{0.38\linewidth}
    \centering
    \includegraphics[width=\linewidth]{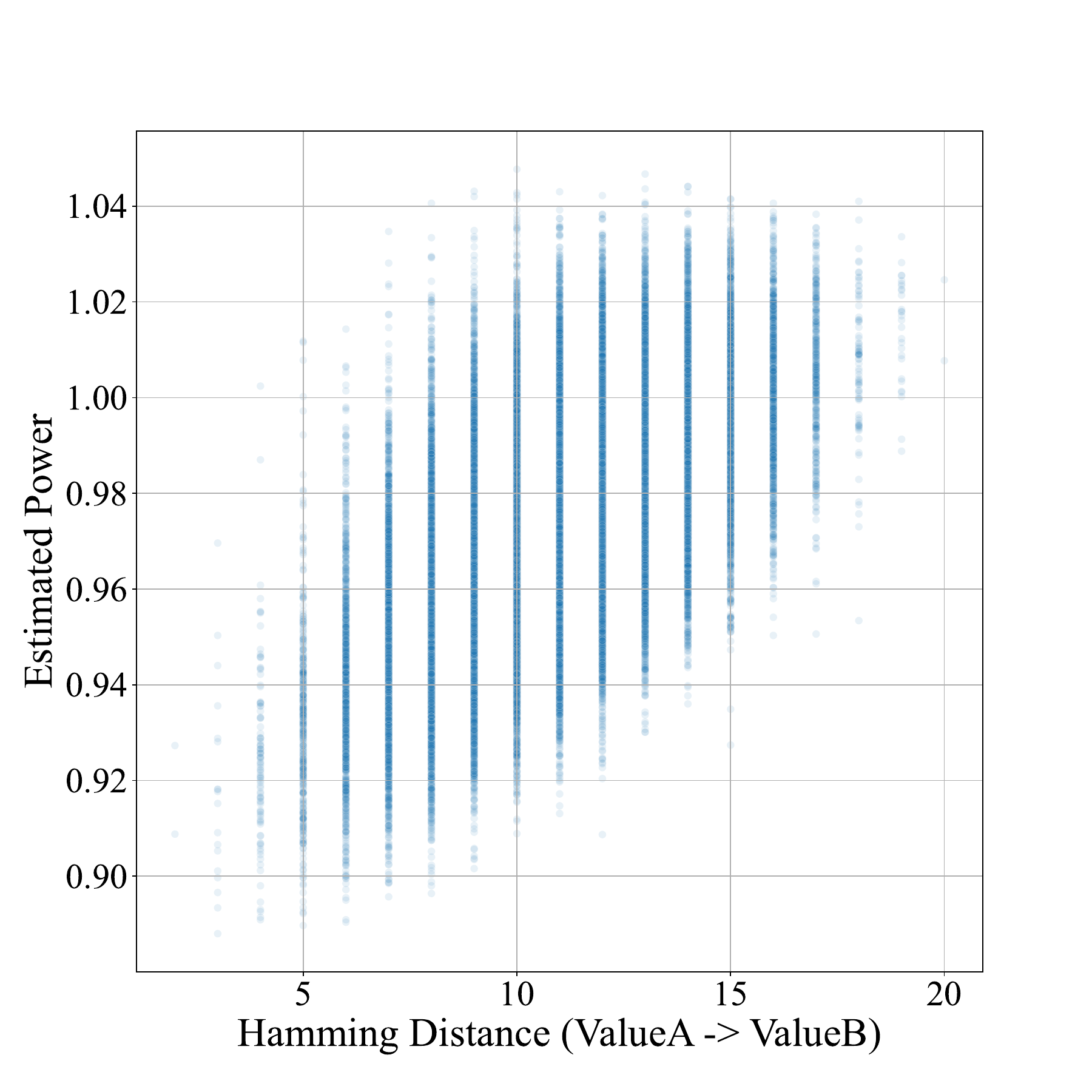}
    \caption{Power vs. Hamming Distance of Transition.}
    \label{fig:hd}
  \end{subfigure}
     \hspace{1.2cm}
  \begin{subfigure}{0.38\linewidth}
    \centering
    \includegraphics[width=\linewidth]{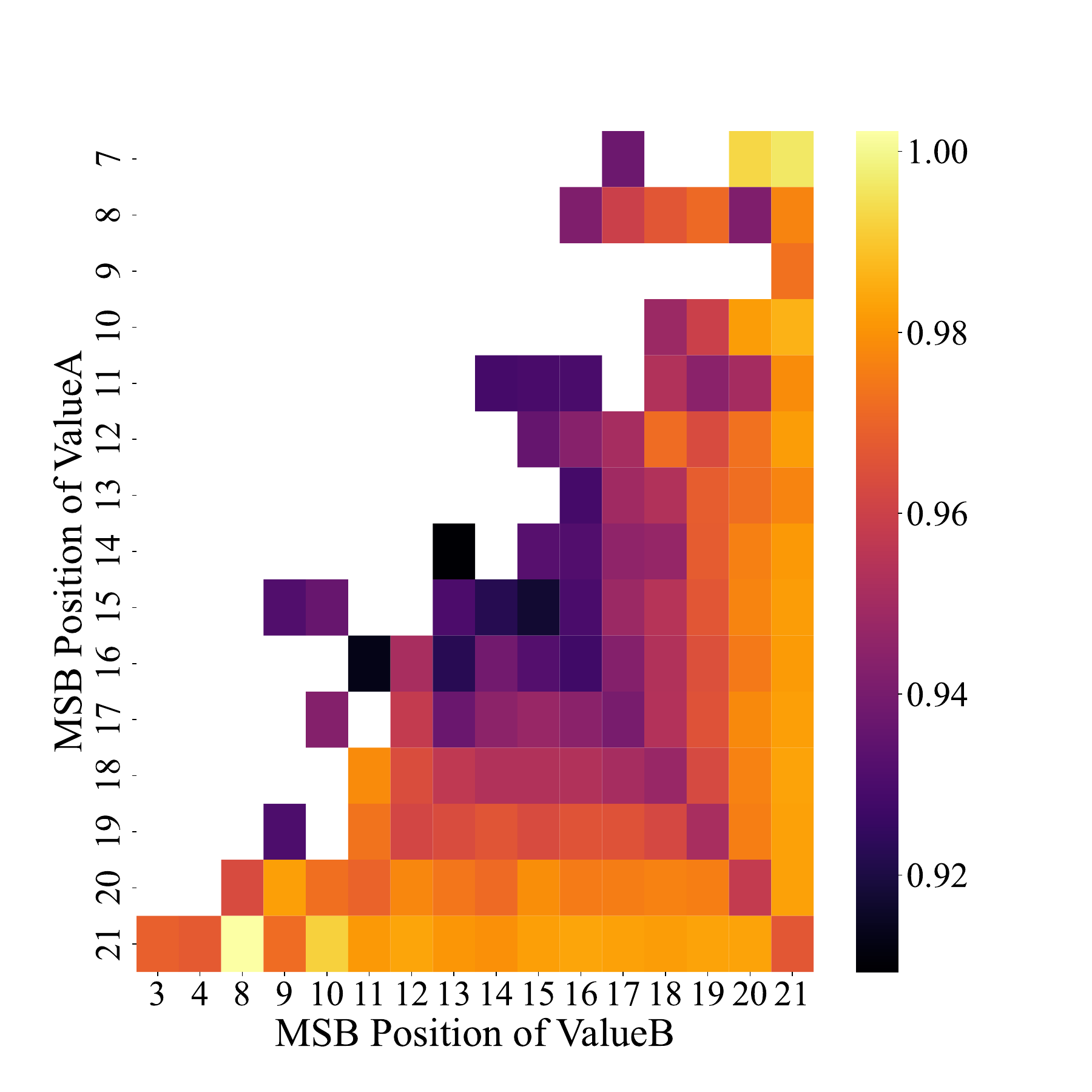}
    \caption{Average power of transitions of different MSB.}
    \label{fig:msb}
  \end{subfigure}
  \vspace{-0.1in}
  \caption{Analysis of how are two grouping metrics rated to power of different transitions.}
  \vspace{-0.1in}
  \label{fig:transition}
\end{figure*}

To design these groups, we examine two bit-level features known to influence switching energy: (i) the most significant bit (MSB) position, which captures high-order carry propagation, and (ii) the Hamming distance (HD) between consecutive values, which reflects the total number of toggled bits.

These features were validated empirically. We generated random partial-sum transitions and measured MAC power under fixed weights. Figure \ref{fig:hd} shows that MAC power increases approximately monotonically with HD. Figure \ref{fig:msb} shows that transitions between partial sums with similar MSB positions consume less energy (visible along the diagonal), while transitions involving higher MSBs exhibit much higher power. These findings indicate that MSB and HD jointly capture the dominant contributors to dynamic switching behavior.
%
We then build a two-stage grouping scheme:

\begin{enumerate}
    \item \textbf{\textit{Stage 1}: Coarse grouping by MSB} \begin{itemize}
        \item MSB values in the range 0–22 are uniformly partitioned into 10 groups.
        \item This grouping exploits the fact that similar MSB positions lead to similar carry-propagation activity and hence similar power.
    \end{itemize}
    \item \textbf{\textit{Stage 2}: Fine-grained grouping by Hamming weight} \begin{itemize}
        \item Within each MSB group, partial sums are further divided into 5 subgroups according to their Hamming weight.
        \item This ensures that transitions within a subgroup maintain small HD, while transitions across subgroups differ significantly in their switching cost.
    \end{itemize}
\end{enumerate}
This results in $10\times 5=50$ total groups. We evaluate grouping quality using a stability ratio defined as the variance of inter-group means divided by the mean intra-group variance. Uniform partitioning performs consistently well across layers and weights, and is therefore adopted in our model.

During profiling, each observed partial-sum transition is mapped to one of the 50 groups, and we estimate the transition probability distribution across these groups.

\subsubsection{Layer-Specific Activation and Partial-Sum Statistics}
Activation distributions vary markedly across layers, especially after applying nonlinearities such as ReLU. We therefore perform layer-specific statistical sampling. For each convolution layer, we independently collect activation transition frequencies and grouped partial-sum transition frequencies.
Once the two distributions are estimated, we generate transition sequences via probabilistic sampling. These synthesized sequences serve as MAC input traces in gate-level simulations, enabling accurate per-weight MAC power measurements that reflect each layer’s unique data characteristics.
Figure \ref{fig:transition} visualizes the activation transition heatmaps for the first two convolution layers of LeNet-5, illustrating substantial variation in transition patterns across layers.
The resulting per-weight MAC power values are later integrated into the layer-wise energy-prioritized compression algorithm in Section~\ref{sec:compression}.

\begin{figure}[htbp]
  \centering 
  \begin{subfigure}{0.49\linewidth}
    \centering
    \includegraphics[width=\linewidth]{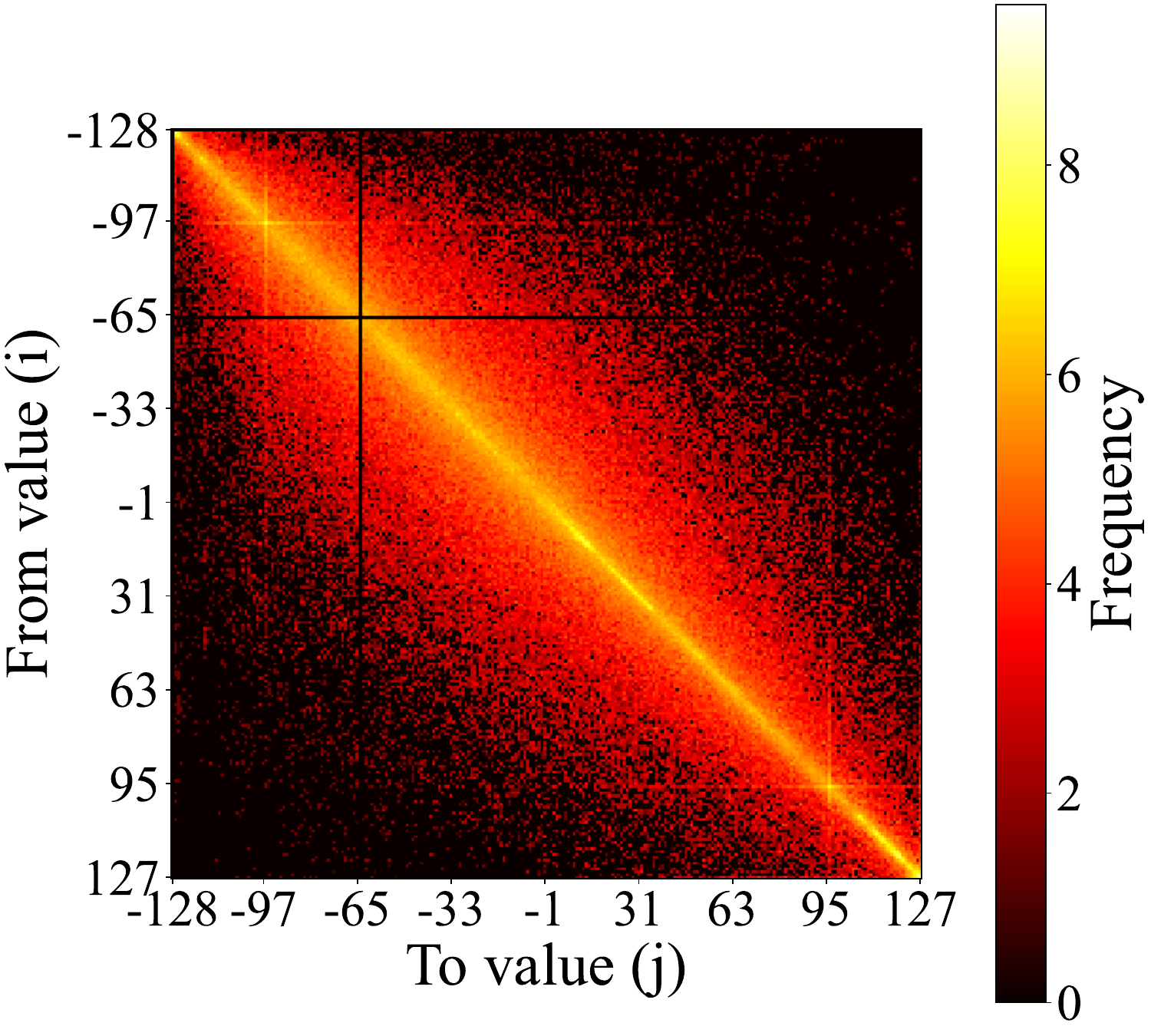}
    \caption{First conv layer }
    \label{fig:act1}
  \end{subfigure}
  \begin{subfigure}{0.49\linewidth}
    \centering
    \includegraphics[width=\linewidth]{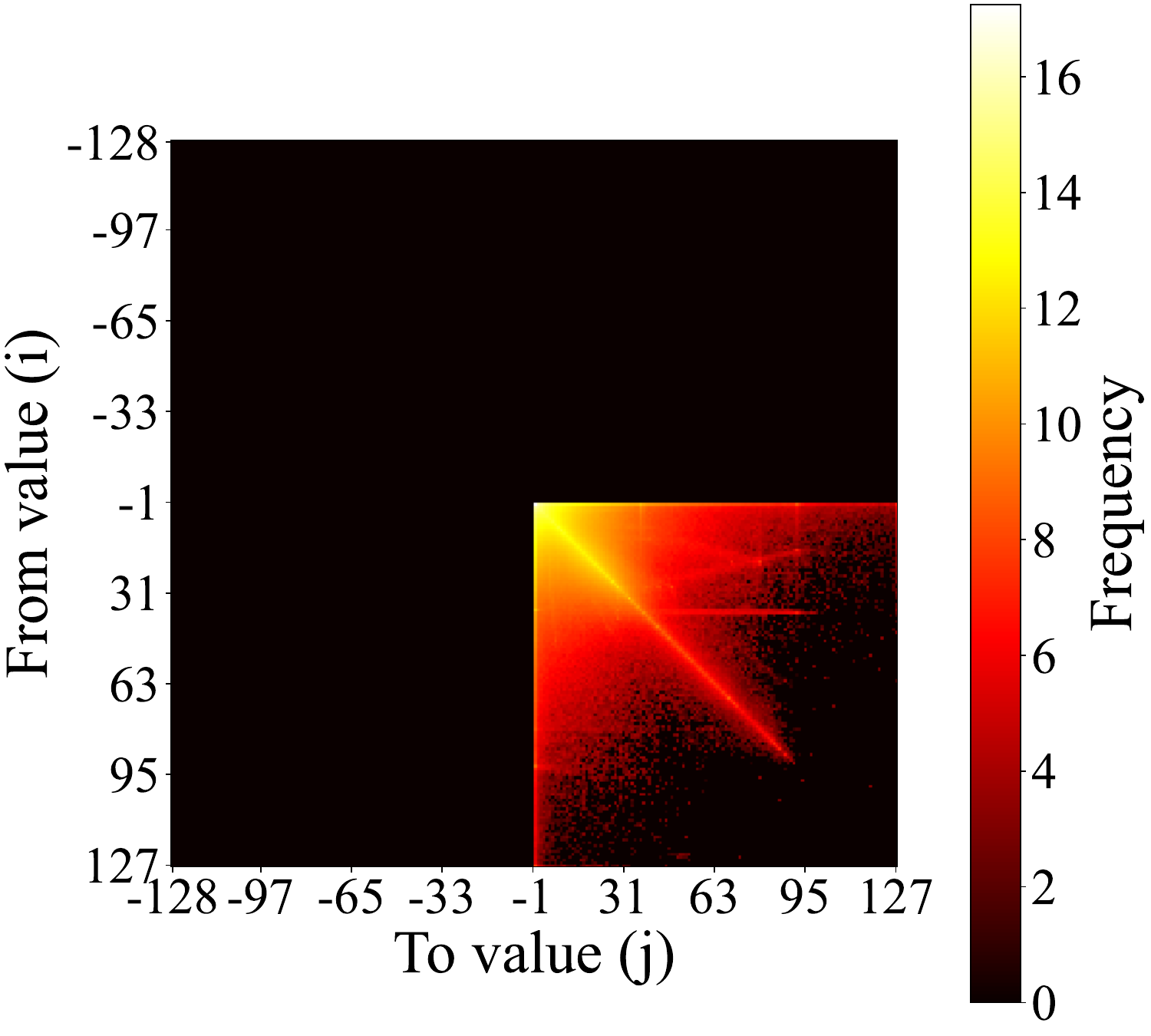}
    \caption{Second conv layer }
    \label{fig:act2}
  \end{subfigure}
  \vspace{-0.1in}
  \caption{Activation transition heatmaps of two convolutional layers of LeNet-5.}
  \vspace{-0.3in}
  \label{fig:transition}
\end{figure}

\subsection{Convolution-Layer Energy Estimation}
\label{sec:energy_modeling_2}
We next estimate the energy consumption of full convolution layers by mapping their computation onto the $64 \times 64$systolic array. We adopt the standard \textit{im2col} transformation to convert each convolution into a matrix multiplication. For an input tensor of shape $(C_{\mathrm{in}}, H, W)$ and filters of shape $(C_{\mathrm{out}}, C_{\mathrm{in}}, k, k)$, we obtain:

$$
W_{\text{mat}} \in \mathbb{R}^{C_{\mathrm{out}} \times (C_{\mathrm{in}} k^2)}, \qquad
X_{\text{col}} \in \mathbb{R}^{(C_{\mathrm{in}} k^2) \times (H_{\mathrm{out}} W_{\mathrm{out}})},
$$
where $H_{\mathrm{out}}$ and $W_{\mathrm{out}}$ denote the spatial dimensions of the output feature maps.

The matrix multiplication $Y = W_{\text{mat}} X_{\text{col}}$ is then partitioned into $64 \times 64$ tiles, each executed by the systolic array. Tiling provides two advantages: (1) it normalizes the simulation environment and eliminates the need to generate layer-specific testbenches; and (2)it mirrors the real hardware scheduling mechanism on fixed-dimension systolic arrays.

For each tile $i$, we simulate the actual weights and activations and record the power $P_{\text{tile}, i}$. The average power of layer $\ell$ is computed as
$$
P_{\text{tile}}^{(\ell)} = \frac{1}{N_\ell} \sum_{i=1}^{N_\ell} P_{\text{tile}, i},
$$
where $N_\ell$ is the number of tiles in layer $\ell$.

In our weight-stationary implementation, each tile requires 128 cycles to complete at clock frequency $f$. We define 
$$
T = 64 \times \frac{1}{f}
$$
and the energy per tile is computed as
$$
E_{\text{tile}} = 2 \cdot P_{\text{tile}}^{(\ell)} \cdot T
$$
Finally, the total energy of layer $\ell$ is
$$
E_\ell = N_\ell \cdot E_{\text{tile}},
$$
which assumes linear accumulation of tile energies, as is typical for weight-stationary systolic arrays without inter-tile data reuse.


\section{Layer-Wise Weight Restriction and Compression}
\label{sec:compression}
In this section, we describe the proposed compression framework that leverages the energy modeling in Section~\ref{sec:energy_modeling} to reduce CNN inference energy. The framework consists of two key components. First, we design a performance–aware weight restriction algorithm that selects a compact set of allowed weight values for each layer, balancing energy reduction and accuracy. Second, we integrate these restricted sets into a layer-wise compression schedule that prioritizes high-energy layers, achieving better energy–accuracy trade-offs under a similar overall compression ratio.

\subsection{Problem Formulation}

Given a trained convolutional network and the MAC-level energy model from Section 3, we assume that, for each layer $\ell$ and each discrete weight value $w$, we can obtain a layer-specific average MAC energy $E_\ell(w)$. Let $\mathcal{W}$ denote the full set of representable weight values (e.g., 8-bit integers). Our goal is to construct, for each layer $\ell$, a restricted weight set $\mathcal{C}_\ell \subset \mathcal{W}$ and a pruning mask such that: (i) the total inference energy is reduced as much as possible; (ii) the validation accuracy remains above a specified threshold; (iii) the overall compression ratio (pruning + quantization) is comparable to or better than existing baselines.
This can be formulated as a constrained optimization problem:
$$
\min_{{\mathcal{C}_\ell},\ \text{masks}} \quad \sum_{\ell} E_\ell(\mathcal{C}_\ell)
\quad \text{s.t.} \quad \text{Acc}({\mathcal{C}_\ell}) \ge \text{Acc}_0 - \delta,
$$
where $E_\ell(\mathcal{C}_\ell)$ denotes the estimated energy of layer $\ell$ when its weights are restricted to $\mathcal{C}_\ell$, $\text{Acc}_0$ is the baseline accuracy, and $\delta$ is an allowable accuracy drop. Solving this problem exactly is intractable; we therefore adopt a greedy strategy that approximates this objective at two levels: within each layer (weight-set selection) and across layers (layer-wise scheduling).

\subsection{Performance–Energy Co-Optimized Weight Set Selection}
\label{sec:compression_1}
A naive approach to weight restriction is to simply select the weight values with the lowest average MAC energy, for example, choosing the 16 lowest-energy weights and forcing the entire network to use only these values. Our empirical results show that although this strategy does reduce energy, it often leads to dramatic accuracy degradation. This indicates that weights with slightly higher energy can still be critical for maintaining model expressiveness.

To address this, we use a two-stage procedure for each layer $\ell$: we first find \textit{a safe starting set} of candidate weights that preserves accuracy (typically of size 32), and then perform \textit{greedy backward elimination} to remove weights that are energy-inefficient but not critical for accuracy.

\subsubsection{Safe Initial Candidate Set}

For each layer $\ell$, we begin by constructing an initial candidate set $\mathcal{C}_\ell^{(0)}$ of size ($K_{\text{init}}$) (e.g., 32). This set can be obtained by combining energy and usage statistics. Concretely, we:
\begin{enumerate}
    \item Rank all weight values $w \in \mathcal{W}$ by a joint score that favors low-energy and frequently used weights in layer $\ell$.
    \item Gradually increase the size of $\mathcal{C}_\ell^{(0)}$ until the network, retrained or fine-tuned with the weights constrained to $\mathcal{C}_\ell^{(0)}$, matches the baseline accuracy within a small tolerance.
\end{enumerate}

This yields a layer-specific set of allowed weights that is already much smaller than the full 8-bit range but does not significantly harm performance.

\subsubsection{Greedy Backward Elimination}

Starting from $\mathcal{C}_\ell^{(0)}$, we aim to reduce its size to a target $K_{\text{target}}$ (e.g., 16) without violating the accuracy constraint. Let $\mathcal{C}_\ell^{(t)}$ denote the candidate set at iteration $t$. At each step, we consider removing one weight value $w \in \mathcal{C}_\ell^{(t)}$ and evaluate its impact on both energy and accuracy.

For each candidate removal $w$, we estimate: (i) the energy saving $\Delta E_\ell(w)$, based on the change in the layer’s energy when $w$ is no longer allowed and all occurrences of $w$ are mapped to the nearest remaining value in $\mathcal{C}_\ell^{(t)} \setminus {w}$;
(ii) the accuracy drop $\Delta \text{Acc}(w)$, measured on a validation set after a short fine-tuning or a calibration pass.
We then define a removal score, for example:
$$
S(w) = \frac{\Delta E_\ell(w)}{\Delta \text{Acc}(w) + \epsilon},
$$
where $\epsilon$ is a small constant for numerical stability. Intuitively, a large $S(w)$ means that removing $w$ yields substantial energy savings at a relatively small accuracy cost.

At each iteration, we: (1) rank all $w \in \mathcal{C}_\ell^{(t)}$ by $S(w)$; (2) select the best candidate $w^\star$ according to this score; (3) tentatively remove $w^\star$ and evaluate the resulting network accuracy. 
If the new accuracy remains above $\text{Acc}_0 - \delta$, we accept the removal and set $\mathcal{C}_\ell^{(t+1)} = \mathcal{C}_\ell^{(t)} \setminus {w^\star}$.
Otherwise, we mark $w^\star$ as \textit{essential} and skip it in subsequent iterations.

The process terminates when either the target size $|\mathcal{C}_\ell^{(t)}| = K_{\text{target}}$ is reached, or when no further removal satisfies the accuracy constraint. In this way, the final candidate set $\mathcal{C}_\ell$ avoids the failure mode of \textit{lowest-energy-only} selection and explicitly balances energy reduction with accuracy preservation.

\subsection{Energy-Prioritized Layer-Wise Compression}
\label{sec:compression_2}

The weight restriction procedure above operates on a single layer. To fully exploit the heterogeneous energy distribution across layers, we integrate it into a layer-wise compression schedule guided by the layer energies derived in Section~\ref{sec:energy_modeling_2}.

Let $E_\ell^{\text{base}}$ denote the estimated energy of layer $\ell$ before compression. We first normalize these values to obtain each layer’s energy contribution:
$$
\rho_\ell = \frac{E_\ell^{\text{base}}}{\sum_j E_j^{\text{base}}}.
$$

Layers with high $\rho_\ell$ contribute more to the total energy and are more promising targets for aggressive compression.
We then sort layers in descending order of $\rho_\ell$ and process them sequentially. For each layer, we consider a set of candidate compression configurations, combining: (1) different pruning ratios (e.g., 0.3, 0.5, 0.7), and (2) different target candidate-set sizes (e.g., 32, 24, 16).
For each configuration of layer $\ell$, we:
(1) apply pruning and weight restriction as described in Section~\ref{sec:compression_2} to obtain the corresponding $\mathcal{C}_\ell$ and mask; (2) estimate the new layer energy $E_\ell$ using the tile-based model from Section~\ref{sec:energy_modeling_2}; (3) evaluate the network accuracy after a short fine-tuning phase.
We then select, for layer $\ell$, the most aggressive configuration that keeps the global validation accuracy above $\text{Acc}_0 - \delta$, and provides a favorable energy saving relative to previously processed layers.

This procedure is repeated for subsequent layers, but compression strength is naturally reduced for layers with smaller $\rho_\ell$, since aggressive compression on low-energy layers yields limited global benefit and may cause unnecessary accuracy loss. As a result, high-energy layers are compressed more aggressively, while low-energy layers are treated more conservatively. Under a similar overall compression ratio, this energy-aware scheduling consistently achieves better energy–accuracy trade-offs than uniform or model-wise compression strategies.

\begin{table}[t]
  \centering
  \small
  \caption{Comparison of proposed method and baselines.}
  \vspace{-0.1in}
  \label{tab:result_all}
  \begin{tabular}{cccc}
    \toprule
    \multirow{2}{*}{\textbf{Network-Dataset}} & \multirow{2}{*}{\textbf{Accuracy}} & \textbf{Energy} & \textbf{Selected} \\
     & & \textbf{Saving} & \textbf{Weights} \\
    \midrule
    LeNet-5-CIFAR-10 (origin) & 78.9\% & -\% & 256 \\
    LeNet-5-CIFAR-10 (~\cite{petri2023powerpruning})  & 78.4\% & 46.0\% & 32 \\
    LeNet-5-CIFAR-10 (Ours)  & 77.8\% & 53.3\% & 16 \\
    \midrule
    ResNet-20-CIFAR-10 (origin) & 92.5\% & -\% & 256 \\
    ResNet-20-CIFAR-10 (~\cite{petri2023powerpruning})  & 88.9\% & 50.9\% & 32 \\
    ResNet-20-CIFAR-10 (Ours)  & 89.4\% & 58.6\% & 16 \\
    \midrule
    ResNet-50-CIFAR-100 (origin) & 81.7\% & -\% & 256 \\
    ResNet-50-CIFAR-100 (~\cite{petri2023powerpruning})  & 78.4\% & 45.3\% & 32 \\
    ResNet-50-CIFAR-100 (Ours)  & 79.6\% & 50.5\% & 16 \\
    \bottomrule
  \end{tabular}
\end{table}

\section{Experimental Results}
\label{sec:experiments}
This section evaluates the effectiveness of the proposed energy-aware, layer-wise compression framework. We begin by reporting overall energy savings and accuracy across several CNN architectures. We then present a detailed layer-wise analysis using ResNet-20, followed by ablation studies comparing our method with global compression strategies and examining the impact of the weight-selection algorithm.

\subsection{Energy Savings}
We evaluate the proposed method on multiple neural networks and datasets, summarized in Table \ref{tab:result_all}. All models are first trained using quantization-aware training, with both weights and activations quantized to 8-bit precision. Using the MAC-level and layer-level energy models developed in Section~\ref{sec:energy_modeling}, we profile each convolution layer to obtain its baseline energy consumption. These energy estimates are then used to configure the weight-selection algorithm and determine the processing order in the layer-wise compression schedule introduced in Section~\ref{sec:compression}.

During compression, pruning is applied first, followed by several epochs of fine-tuning and the introduction of weight set constraints. All experiments are conducted on a single NVIDIA RTX 3060 GPU. Energy consumption is measured by simulating convolution-layer execution on a $64\times 64$ weight-stationary systolic array in Modelsim and evaluating switching power using Synopsys Design Compiler with the NanGate 15 nm library~\cite{silicon15} at 5 GHz.

Table \ref{tab:result_all} presents the resulting energy savings, accuracy changes, and number of selected weights. Across all models, our method achieves substantial reductions in convolution-layer energy with minimal losses in accuracy. Compared to existing approaches~\cite{petri2023powerpruning}, our method achieves higher energy savings even with smaller weight sets, owing to the layer-wise compression strategy and the fine-grained energy modeling.

\subsection{Layer-Wise Energy Reduction}

Table \ref{tab:result_lw} reports the layer-wise reductions on ResNet-20 (CIFAR-10). We group convolutional layers according to their BasicBlock identity because pruning primarily affects these blocks. Using the layer-level energy model from Section~\ref{sec:compression}, we identify Block 2 and Block 4 as the most energy-intensive regions of the network. These blocks are therefore compressed first using more aggressive pruning levels and smaller weight sets.

As shown in Table \ref{tab:result_lw}, early processing of high-energy layers yields substantial energy savings. Layers with smaller energy contributions (e.g., Block 6 and Block 9) are processed afterward with milder compression, preserving model accuracy while still contributing to overall energy reduction. This schedule aligns with the energy-prioritized strategy described in Section 4 and directly explains the strong energy–accuracy trade-offs achieved in Table \ref{tab:result_all}.

\begin{table}[t]
  \centering
  \small
  \caption{Layer-wise energy saving in ResNet-20-CIFAR-10}
  \label{tab:result_lw}
  \begin{tabular}{ccccc}
    \toprule
    \multirow{2}{*}{\textbf{Layer}} & \textbf{Prune} & \textbf{Selected} & \textbf{Energy} & \multirow{2}{*}{\textbf{Share}}\\
     & \textbf{Ratio} & \textbf{Weights} & \textbf{Saving} & \\
    \midrule
    Block 2 (Conv3, Conv4) & 0.7 & 16 & 61.8\% & 21.1\% \\
    Block 4 (Conv7, Conv8) & 0.75 & 16 & 63.2\% & 23.7\%\\
    Block 6 (Conv11, Conv12) & 0.5 & 16 & 51.2\% & 7.6\%\\
    Block 9 (Conv17, Conv18) & 0.5 & 16 & 48.3\% & 3.9\%\\
    \bottomrule
    \vspace{-0.2in}
  \end{tabular}
\end{table}


\subsection{Ablation Studies}

\subsubsection{Layer-wise vs. Global Compression}

To isolate the effect of the layer-wise strategy, we compare it against a traditional global compression approach under matched pruning ratios and weight-set cardinalities. Results are shown in Table \ref{tab:results_lvg}.

For both moderate (32-weight) and aggressive (16-weight) constraints, the layer-wise approach consistently yields higher energy savings and significantly better accuracy. In particular, when the weight set is reduced to 16 values, the global method suffers severe accuracy degradation, while the layer-wise strategy maintains accuracy close to the baseline. These results highlight the importance of prioritizing high-energy layers as described in Section~\ref{sec:compression_2}.

\begin{table}[t]
  \centering
  \small
  \caption{Comparison of Layer-wise vs. Global Strategies on ResNet-20}
  \label{tab:results_lvg}
  \begin{tabular}{ccccc}
    \toprule
    \multirow{2}{*}{\textbf{Method}} & \textbf{Prune} & \textbf{Selected} & \textbf{Energy} & \multirow{2}{*}{\textbf{Accuracy}} \\
     & \textbf{Ratio} & \textbf{Weights} & \textbf{Saving} & \\
    \midrule
    Block-4-Global & 0.5 & 32 & 46.8\% & 89.1\% \\
    Block-4-Layer-wise & 0.5 & 32 & 48.3\% & 89.8\% \\
    \midrule
    Block-4-Global & 0.5 & 16 & 50.1\% & 82.0\% \\
    Block-4-Layer-wise & 0.5 & 16 & 51.8\% & 89.4\% \\
    \midrule
    Block-2-Global & 0.7 & 32 & 52.8\% & 86.3\% \\
    Block-2-Layer-wise & 0.7 & 32 & 54.3\% & 88.6\% \\
    \bottomrule
    \vspace{-0.2in}
  \end{tabular}
\end{table}


\begin{table}[htbp]
  \centering
  \caption{Effectiveness of the Weight Selection Algorithm on ResNet-20}
  \label{tab:results_selec}
  \begin{tabular}{ccc}
    \toprule
     \textbf{Selected Weights} & \textbf{Energy Saving} & \textbf{Accuracy} \\
    \midrule
     Naive (Top 16) & 59.3\% & 59.6\%  \\
     Naive (Top 20) & 57.5\% & 89.6\%  \\
     Optimized (Selected 16) & 58.6\% & 89.4\% \\
    \bottomrule
    \vspace{-0.2in}
  \end{tabular}
\end{table}

\subsubsection{Impact of Different Compression Components}

\begin{figure}[t]
  \centering 
  \begin{subfigure}{0.43\linewidth}
    \centering
    \includegraphics[width=\linewidth]{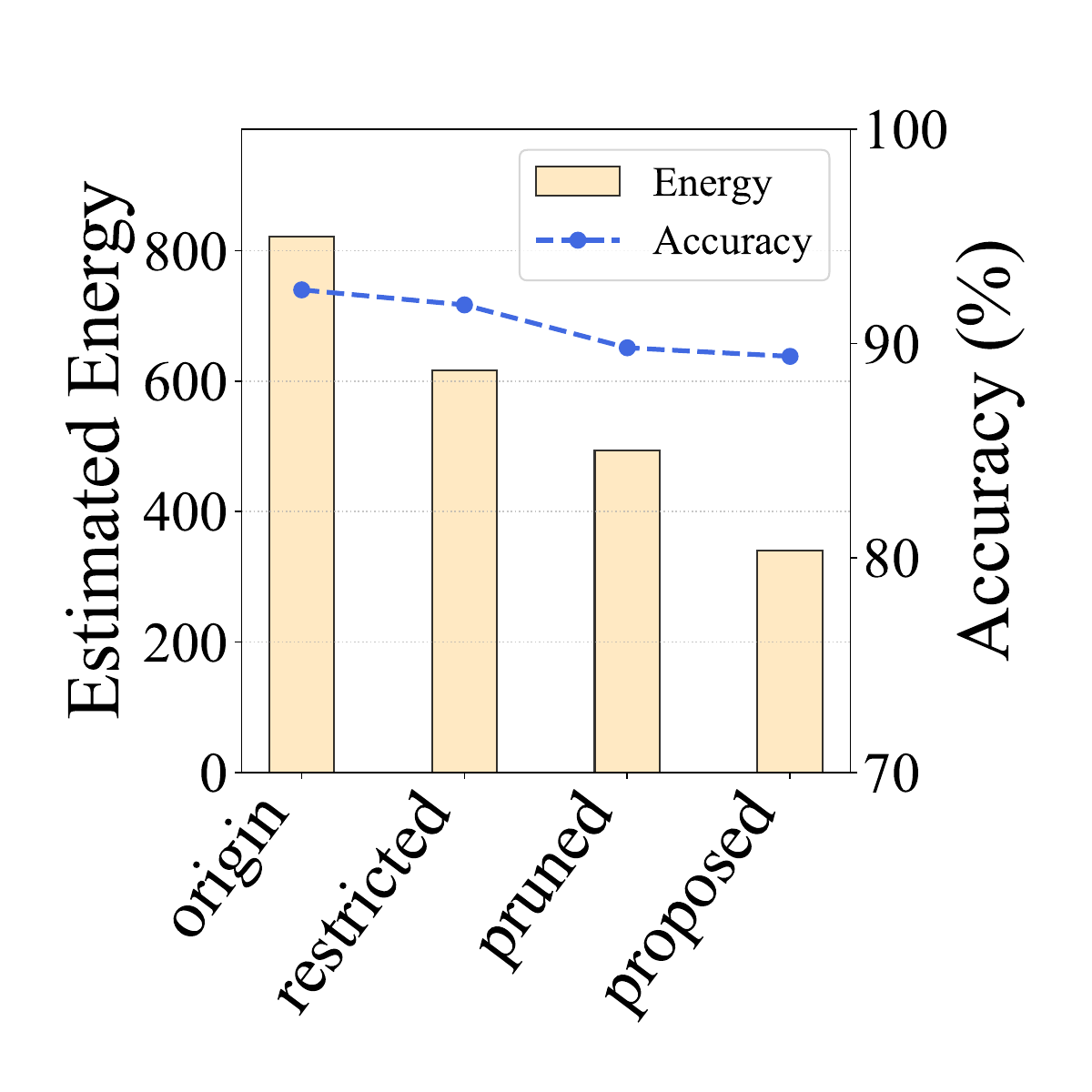}
    \caption{Overall model.}
    \label{fig:hd}
  \end{subfigure}
  \begin{subfigure}{0.56\linewidth}
    \centering
    \includegraphics[width=\linewidth]{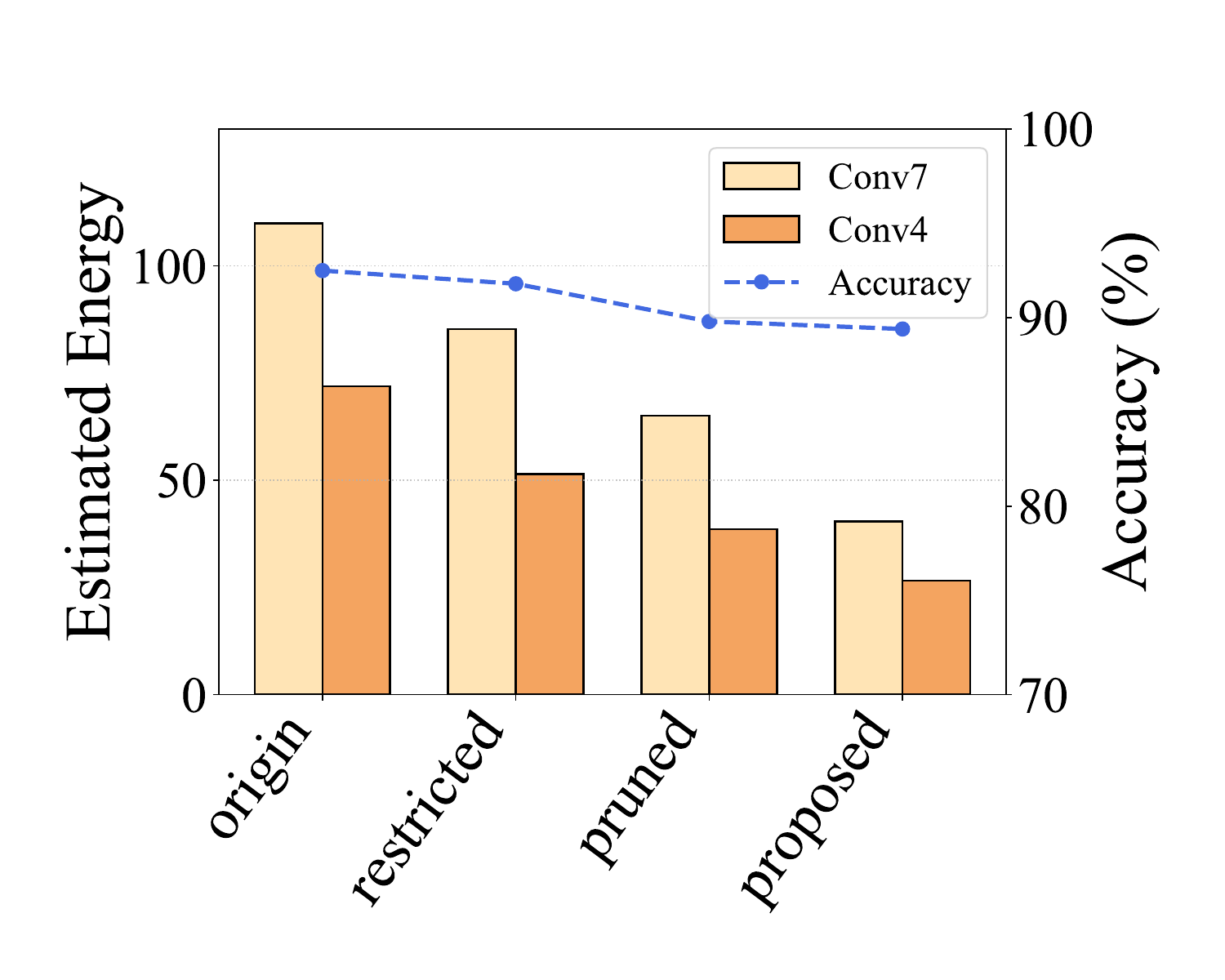}
    \caption{Two conv layers.}
    \label{fig:msb}
  \end{subfigure}
  \vspace{-0.2in}
  \caption{Comparison of different compression methods on ResNet-20.}
  \vspace{-0.2in}
  \label{fig:abla}
\end{figure}

Figure \ref{fig:abla} compares the effects of pruning, weight restriction, and their combination. Both techniques independently contribute to energy savings; however, applying them together achieves substantially larger reductions, confirming that pruning and weight restriction provide complementary benefits.

\subsubsection{Effectiveness of the Weight-Selection Algorithm}
Table \ref{tab:results_selec} compares our weight-selection algorithm against a naive approach that selects the 16 weights with the lowest MAC energy.

The naive strategy results in catastrophic accuracy degradation, highlighting the necessity of balancing energy and expressiveness. Even the best naive configuration (top 20 lowest-energy weights) cannot match the accuracy of our optimized 16-weight configuration. This confirms the effectiveness of the greedy elimination strategy introduced in Section~\ref{sec:compression_1}.

\section{Conclusion}
\label{sec:conclusion}
We presented an energy-aware, layer-wise compression framework for CNNs on systolic-array accelerators. By combining layer-specific activation statistics with a compact MSB–Hamming-distance grouping scheme, our modeling yields accurate MAC-level and layer-level energy estimates. Leveraging this model, we proposed a two-stage compression method that couples pruning with an energy–accuracy co-optimized weight-selection algorithm, and a layer-wise schedule that prioritizes high-energy layers.
Experiments on multiple networks demonstrate that the proposed approach achieves higher energy savings than prior work with minimal accuracy loss, especially under aggressive weight constraints. Across all evaluated models, the framework consistently achieves 50–60\% energy reduction within 3\% accuracy loss. These results highlight the effectiveness of detailed energy modeling and layer-wise optimization for improving the efficiency of CNN inference on modern accelerators.

\bibliographystyle{ACM-Reference-Format}
\bibliography{sample-base}










\end{document}